\documentclass[aps,prl,twocolumn,showpacs]{revtex4}
\usepackage{amsmath}    
\usepackage{graphicx}   
\usepackage{verbatim}   
\usepackage{color}      
\usepackage{subfigure}  
\usepackage{hyperref}   

\usepackage{amsmath}

\hypersetup{pdfborder = {0 0 0},colorlinks=true}
\renewcommand{\th}{\operatorname{tanh}}
\newcommand{\arcth}{\operatorname{arctanh}}

\renewcommand{\H}{{\mathcal H}}
\newcommand{\s}{{\bf s}}
\newcommand{\plus}{^{\,+}}
\newcommand{\minus}{^{\,-}}
\newcommand{\var}{\operatorname{var}}

\begin{document}
\title{Mean-field theory for the inverse Ising problem at low temperatures}

\author{H. Chau Nguyen and Johannes Berg}
\email{cnguyen@thp.uni-koeln.de and berg@thp.uni-koeln.de}
\affiliation{University of Cologne, Institute for Theoretical Physics\\Z\"{u}lpicher Stra{\ss}e 77, 50937 K\"{o}ln, Germany}

\begin{abstract} 
The large amounts of data from molecular biology and neuroscience have lead to a renewed interest in the inverse Ising problem: how to reconstruct parameters of the Ising model (couplings between spins and external fields) from a number of spin configurations sampled from the Boltzmann measure. To invert the relationship between model parameters and observables (magnetisations and correlations) mean-field approximations are often used, allowing to determine model parameters from data. However, all known mean-field methods fail at low temperatures with the emergence of multiple thermodynamic states. Here we show how clustering spin configurations can approximate these thermodynamic states, and how mean-field methods applied to thermodynamic states allow an efficient reconstruction of Ising models also at low temperatures. 
\end{abstract}


\maketitle

Taking a set of spin configurations sampled from the equilibrium distribution of an Ising model, can the underlying couplings between spins be reconstructed from a large number of such samples? This inverse Ising problem is a paradigmatic inverse problem with applications in neural biology~\cite{bialek,monasson}, protein structure determination~\cite{weigt}, and gene expression analysis~\cite{braunstein}. Typically a large number of spins (representing the states of neurons, genetic loci, or genes) is involved, as well as a large number of interactions between them.

Such large system sizes makes the inverse Ising model intrinsically difficult: Solving the inverse problem involves first solving the Ising model, in some manner, for a given set of couplings and external fields. Then one can ask how couplings between spins and external fields need to be adjusted in order to match the inferred model with the observed statistics of the samples. An early and fundamental approach to the inverse Ising model, Boltzmann machine learning ~\cite{ackley1985}, follows this prescription quite literally. Proceeding iteratively, couplings and fields are updated in proportion to the remaining differences of magnetisations and two-point-correlations resulting from the current model parameters and the corresponding values observed in data. To compute the magnetisations and two-point-correlations, each iteration involves a numerical simulation of the Ising model, so this approach is limited to small systems. 

Instead, mean-field theory is the basis of many approaches to the inverse Ising problem used in practice~\cite{kappen,Roudi_FCN}. Under the mean-field approximation, the Ising model can be solved easily for the magnetisations and correlations between spins. The mean-field solution is then inverted (see below) to yield the parameters of the model (couplings and external fields) as a function of the empirical observables (magnetisations and correlations). Yet, as temperature is decreased and correlations between spins grow and become more discernible, the reconstruction given by mean-field theory becomes less accurate, not as one might expect, more accurate. This effect has been called ``an embarrassment to statistical physics''~\cite{aurell_priv}. Mean-field reconstruction of the Ising model even breaks down entirely near the transition to a low-temperature phase~\cite{Mezard_JOP}: in the low-temperature phase there is no correlation between reconstructed and underlying couplings. This low-temperature failure equally affects all refinements related to mean-field theory like the TAP approach~\cite{kappen,Mezard_JOP,Roudi_FCN}, susceptibility propagation~\cite{Welling_NC,Mezard_JOP}, the Sessak--Monasson expansion~\cite{Sessak}, and Bethe reconstruction~\cite{NguyenBerg}.

The breakdown of mean-field reconstructions can have different roots: the emergence of multiple thermodynamic states at a phase transition, an increasing correlation length at lower temperatures, or the freezing of the spins into a reduced set of configurations at low temperatures requiring more samples to measure the correlations between spins. To address this issue, we first consider a very simple case where mean-field theory is exact: the Curie--Weiss model.  The zero-field Hamiltonian of $N$ binary spins $s_i$ is $\H_J(\s)=-J/N \sum_{i<j} s_i s_j$ with $J=1$. This corresponds to equal couplings $J^0_{ij}=J/N$ between all pairs of spins, a fact that is of course not known when reconstructing the couplings.  $M$ samples of spin configurations are taken from the equilibrium measure $\exp\{-\beta \H_J(\s)\}/Z$, where $\beta$ is the inverse temperature and $Z$ is the partition function. In a real-life reconstruction, these configurations would come from experimental measurements of neural activity, gene expression levels, etc. One then can calculate the observed magnetisations $\bar{m}_i=\frac{1}{M} \sum_{\mu} s^{\mu}_i$ and connected correlations $\bar{c}_{ik}= \frac{1}{M} \sum_{\mu} s^{\mu}_i s^{\mu}_k - \bar{m}_i \bar{m}_k$, with $\mu =1,\ldots,M$ denoting the sampled configurations.

The mean-field prediction for the magnetisations of the Curie-Weiss model is given by the solution of the self-consistent equation
\begin{equation}
\label{nMF_selfconsistent}
m_i=\th ( \sum_{j \neq i} J_{ij} m_j + h_i) \ ,
\end{equation}
where the couplings are rescaled with temperature $J_{ij} = \beta J^0_{ij}$. The connected correlations follow from (\ref{nMF_selfconsistent}) by considering the linear response
\begin{eqnarray}
\label{nMF_correlations}
c_{ik}
&=& \frac{\partial m_i}{\partial h_k} = (1-m_i^2) \left(  \sum_{j \neq i} J_{ij} \frac{\partial m_j}{\partial h_k} + \delta_{ik} \right) \nonumber\\
&=&(1-m_i^2) \left(  \sum_{j \neq i} J_{ij} c_{jk} + \delta_{ik} \right)
 \ .
\end{eqnarray}

Inserting the observed magnetisations and correlations into (\ref{nMF_correlations}) gives~\cite{kappen} 
\begin{equation}
\label{nMF_reconstruction}
\sum_{j \neq i} J_{ij} \bar{c}_{jk}=  - \delta_{ik} + \bar{c}_{ik}/(1-\bar{m}_i^2) \ ,
\end{equation}
which can be solved directly for the couplings $J_{ij} = - (\bar{c}^{-1})_{ij}$ ($i \neq j$) and the fields $h_i=\arcth \bar{m}_i - \sum_{j \neq i} J_{ij} \bar{m}_j$ using  (\ref{nMF_selfconsistent}).

\begin{figure}[b!]
\includegraphics[width = .5\textwidth]{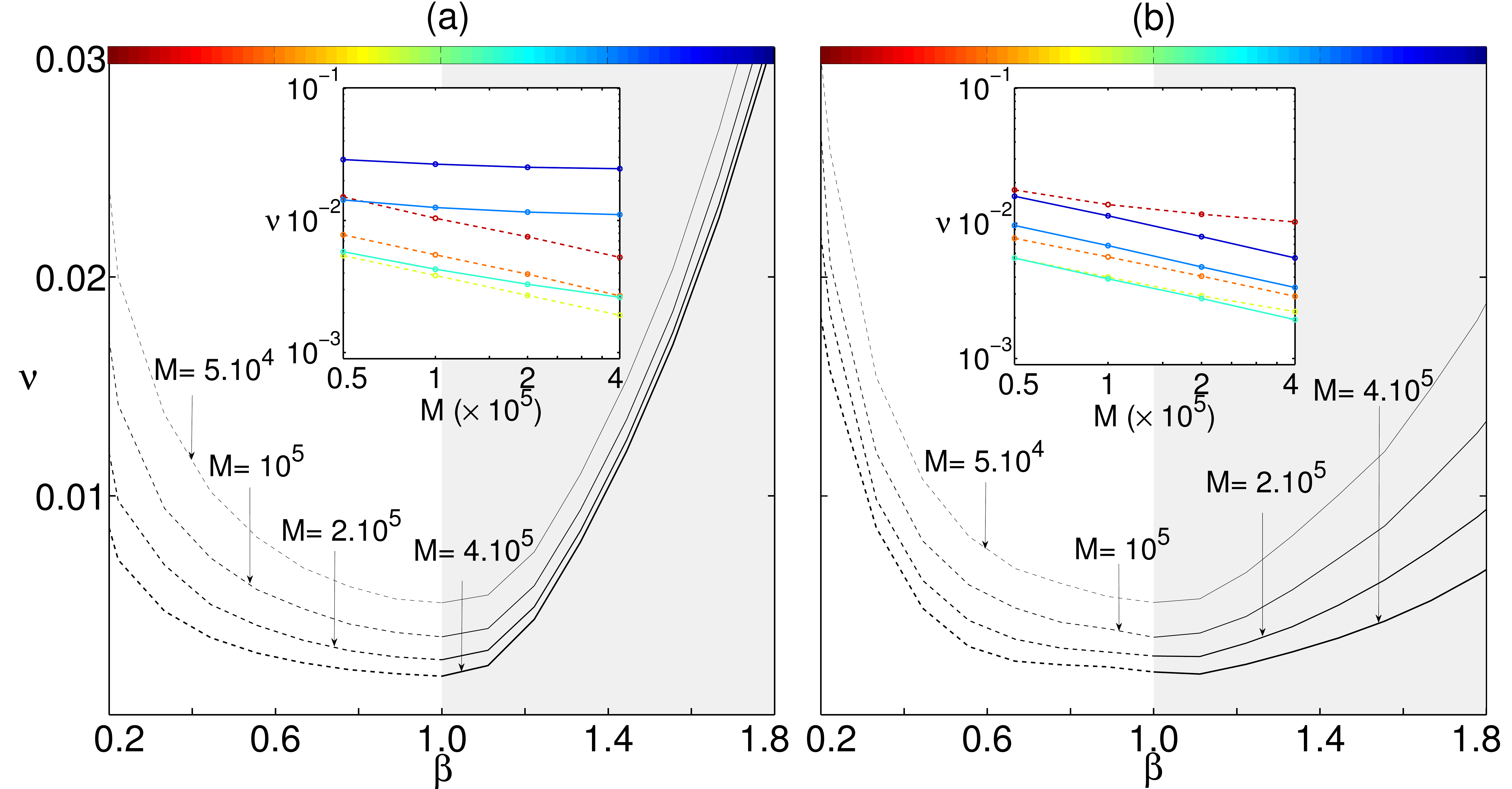}
\caption{{\bf Reconstructing couplings of the Curie--Weiss ferromagnet.} The root-mean-squared deviation between the reconstructed couplings and underlying couplings, $\nu = \sqrt{ \frac{2}{N(N-1)} \sum_{i<j} (J_{ij}/\beta- J/N)^2}$, is plotted against the 
inverse temperature $\beta$ for different numbers of configurations ($J=1$). The system size is $N=100$. The insets show this deviation on a logarithmic scale versus the number of samples $M$ at different inverse temperatures indicated by the colors of the curves ($\beta=0.3,0.58,0.86,1.14,1.42,1.7$). {\bf a)} Reconstruction based on a single thermodynamic state breaks down in the low-temperature phase $\beta>1$ and the deviation between reconstructed and underlying couplings does not vanish with an increasing number of samples $M$ (see the blue curves in the inset). {\bf b)} Reconstruction based on two thermodynamic states is asymptotically exact. The deviation between reconstructed and underlying couplings vanishes as $M^{-1/2}$ at low temperatures.
\label{fig_CW}}
\end{figure}

Figure~\ref{fig_CW}a shows how well this reconstruction performs at different inverse temperatures $\beta$ and different number of samples $M$. For $\beta<\beta_c=1$, the reconstruction error goes to zero with the number of samples as $M^{-1/2}$: since for the Curie--Weiss model the self-consistent equation~(\ref{nMF_selfconsistent}) is exact, the reconstruction is limited only by fluctuations of the measured correlations resulting from the finite number of samples and by the finite system size.  

Yet for $\beta>\beta_c$, the difference between the underlying couplings and the reconstructed couplings does not vanish with increasing number of samples. While the self-consistent equation~(\ref{nMF_selfconsistent}) is still correct, the identification of its solutions 
with the observed magnetisations $\bar{m}_i$ is mistaken. For the ferromagnetic phase at $\beta>\beta_c$, there are two solutions of the self-consistent equation, denoted $m_i^{\pm}=\pm m$. The observed magnetisations are averages over these two thermodynamic states and they have nothing to do with either of the two solutions of~(\ref{nMF_selfconsistent}). The same holds for the connected correlations $c_{ij}^{+}$ and $c_{ij}^{-}$ in the two states, and the observed correlations $\bar{c}_{ij}$.   
Any method explicitly or implicitly connecting the magnetisation in low temperature states with the average magnetisation over samples will thus fail at low temperatures. 
Note that this does not affect Boltzmann machine learning, where the magnetisation is averaged over all states. 

A simple cure suggests itself: Since each sample stems from one of the two thermodynamic states, we divide the $M$ configurations into those configurations with positive total magnetisation $\sum_i s_i^{\mu}$, and those with negative total magnetisation.  Then the magnetisations in the two thermodynamic states can be calculated separately, giving $\bar{m}_i^+=\frac{1}{M_+} \sum_{\mu \in +} s^{\mu}_i$ and similarly for $\bar{m}_i^-$ and the connected correlations. Identifying these magnetisations with the solutions of the self-consistent equation~(\ref{nMF_selfconsistent}), we obtain in the place of~(\ref{nMF_reconstruction}) \textit{two} sets of equations
\begin{eqnarray}
\sum_{j \neq i} J_{ij}  \bar{c}_{jk}\plus&=&   - \delta_{ik} + \bar{c}_{ik}\plus/(1-(\bar{m}_i^+)^2) \label{MF_reconstruction_up}  \ , \\
\sum_{j \neq i} J_{ij}  \bar{c}_{jk}\minus &=&   - \delta_{ik} + \bar{c}_{ik}\minus/(1-(\bar{m}_i^-)^2) \label{MF_reconstruction_down} \ .
\end{eqnarray}
Reconstructing the couplings using a single state only, by solving say~(\ref{MF_reconstruction_up}), the observed positive magnetisation can be accounted for equally well by positive external fields (even though the samples were generated by a model with zero field), or alternatively, by ferromagnetic couplings between the spins. One finds that solving~(\ref{MF_reconstruction_up}) leads to an  underestimate of the couplings, and \textit{positive} external fields calculated by~(\ref{nMF_selfconsistent}) follow. Correspondingly, basing the reconstruction only on data from the down state by solving~(\ref{MF_reconstruction_down}) also leads to an underestimate of the couplings, and large \textit{negative} fields. This effect has already been noted in the context of the inverse Hopfield problem~\cite{zecchina}. We thus demand that the reconstructed fields obtained from either state are equal to each other
\begin{equation}
\label{MF_fields}
\sum_{j \neq i} J_{ij} (\bar{m}_j^+ - \bar{m}_j^-) = \arcth \bar{m}_i^+ - \arcth \bar{m}_i^- 
\end{equation}
and claim that jointly solving equations~\eqref{MF_reconstruction_up}, \eqref{MF_reconstruction_down}, and \eqref{MF_fields} gives the correct mean-field reconstruction at low temperatures. 

Already equations ~\eqref{MF_reconstruction_up} and \eqref{MF_reconstruction_down} are two linear equations per coupling variable, so in general there is no solution to these equations. However, we expect that the underlying couplings used to generate the $M$ configurations actually solve these equations, at least up to fluctuations due to the finite number of configurations sampled and the finite size effect. For an overdetermined linear equation of the form ${\bf A \cdot  x}= {\bf b}$ with vectors of different lengths ${\bf x}$ and ${\bf b}$ and a non-square matrix ${\bf A}$, the Moore--Penrose pseudoinverse ${\bf A^+}$~\cite{moore,penrose} gives a 
least-square solution ${\bf x}={\bf A^+ \cdot  b}$ such that the Euclidean  norm $|| {\bf A\cdot  x} - {\bf b}||_2$ is minimized. In this sense, the Moore--Penrose pseudoinverse allows to solve~\eqref{MF_reconstruction_up}, \eqref{MF_reconstruction_down}, and \eqref{MF_fields} as well as possible. The linear equations~\eqref{MF_reconstruction_up}, \eqref{MF_reconstruction_down}, and \eqref{MF_fields} can be written as a single matrix equation ${\bf J \cdot A}={\bf B}$, where ${\bf A}$ is the $N \times (2N+1)$ matrix $\left( {\bf \bar{c}^+, \bar{c}^-, \bar{m}^+ - \bar{m}^- }\right) $ and ${\bf B}$ is the $N \times (2N+1)$ matrix $( {\bf \bar{b}^+, \bar{b}^-, \tilde{m}^+ - \tilde{m}^-) }$, with $\bar{b}_{ij}^+= -\delta_{ij} + \bar{c}_{ij}\plus/(1-(\bar{m}_i^+)^2)$ and analogously for $\bar{b}_{ij}^-$, and $\tilde{m}_i^+= \arcth \bar{m}_i^+$ and analogously for $\tilde{m}_i^-$. The Moore--Penrose inverse is calculated using singular value decomposition~\cite{nr} and right-multiplied with ${\bf B}$ to obtain the the optimal solution ${\bf J}$. In general, this matrix will not be symmetric, and we use $(J_{ij}+J_{ji})/2,\ i\neq j$ for the reconstructed couplings. The external fields can be computed for each state from $h_i^+=\arcth \bar{m}_i^+ - \sum_{j \neq i} J_{ij} \bar{m}_j^+$, and analogously for $h_i^-$.  Their average over the two states is used for the reconstructed fields.

Figure~\ref{fig_CW}b shows how the reconstruction error now vanishes as $M^{-1/2}$ also in the ferromagnetic phase, albeit with a prefactor which grows as the temperature decreases. So while the mean-field reconstruction from many samples is still successful at low temperatures, more configurations are needed to obtain a certain reconstruction error: At very low temperatures, most spins will be in the same state (either up or down); the connected correlations are small as a result and require many samples for their accurate determination. The quality of the reconstruction depends on configurations being correctly assigned to the thermodynamic states. Artificially introducing mistakes in this assignment, we find the reconstruction error increases linearly with the fraction of mistakes in the assignment of configurations to states. 

In practice, couplings between spins will not all be equal to each other, like in the Curie--Weiss model. Ferromagnetic as well as antiferromagnetic couplings may be present in magnetic alloys, neurons have excitatory and inhibitory interactions, regulatory interactions between genes can either enhance or suppress the expression of a target gene.  The Curie--Weiss ferromagnet is not a good model for all those cases where the couplings are of different signs and magnitudes. In fact, in models where all spins interact with each other via couplings that can be positive or negative~\cite{SK}, the low-temperature regime may be characterized not by two, but by many thermodynamic states~\cite{TAP,MezardParisiVirasoro}. These so-called glassy states cannot be identified simply by the total magnetisation of each sample, as is the case for the ferromagnet. Nevertheless, configurations $\mu,\mu^{\prime}$ from the same thermodynamic state are typically close to each other, having a large overlap $(1/N) \sum_i s_i^{\mu} s_i^{\mu^{\prime}}$. Glassy thermodynamic states thus appear as clusters in the space of configurations~\cite{domanystaufferhartmann,montanarikrzalaparisi_pnas}.

We use the $k$-means clustering algorithm~\cite{kmeans} to find these clusters in the sampled spin configurations. Starting with a set of randomly chosen and normalized cluster centres, each configuration is assigned to the cluster centre it has the largest overlap with. Then the cluster centres are moved to lie in the direction of the centre of mass of all configurations assigned to that cluster, and the procedure is repeated until convergence. We also tried out different algorithms from the family of hierarchical clustering methods, but found no significant difference in the reconstruction performance.  Then magnetisations and connected correlations are computed for each cluster separately. Equations ~\eqref{MF_reconstruction_up}, \eqref{MF_reconstruction_down}, and \eqref{MF_fields} can be written for $k$ thermodynamic states. The mean-field equation for each state and the condition that the external fields are equal in all states can be written again in the form of a matrix equation ${\bf J \cdot A}={\bf B}$.  ${\bf A}$ is the $N \times (kN+k)$ matrix $\left( {\bf \bar{c}^1, \ldots, \bar{c}^k, \bar{m}^1 - \langle\bar{m} \rangle ,\ldots, \bar{m}^k - \langle\bar{m} \rangle } \right)$ 
where $\langle \cdot \rangle$ denotes the average over clusters, $\langle{\bf \bar{m}} \rangle = (1/k) \sum_{a=1}^k {\bf \bar{m}^a}$, and analogously for  ${\bf B}$. 
The pseudoinverse of $A$ can be computed in ${\mathcal O}(kN^3)$ steps~\cite{nr}, so up to a factor of $k$ coming from the number of clusters, this is just as fast as 
the high-temperature mean-field reconstruction based on Gaussian elimination to invert the correlation matrix. 

\begin{figure}[tbh!]
\includegraphics[width = .5\textwidth]{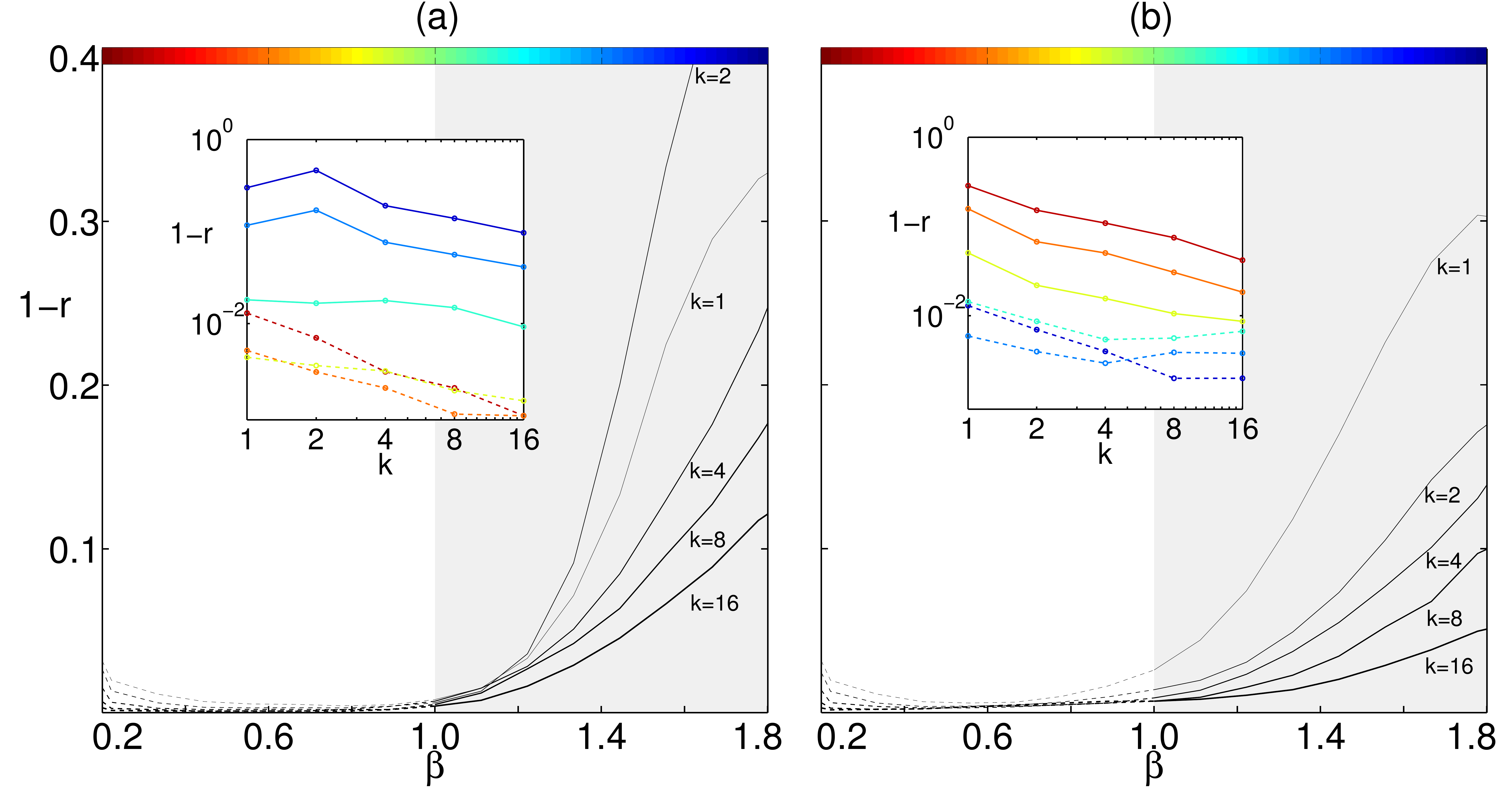}
\caption{{\bf Reconstructing couplings of the Sherrington--Kirkpatrick model.} The Pearson correlation coefficient $r$ quantifies the correlation between reconstructed couplings and underlying couplings,  
$r=\frac{\frac{1}{N(N-1)}\sum_{i \neq j}( J_{ij}-\langle J \rangle) (J^0_{ij}-\langle J^0 \rangle) }{\sqrt{\var( J)\var( J^0)}}$, where $\langle J \rangle$ , $\var(J)$
are the mean  and  variance of the reconstructed couplings across bonds, and similarly for the underlying couplings. $r=1$, or $1-r=0$,  corresponds to perfect reconstruction. The main plots show $1-r$ against the  inverse temperature $\beta$ for different numbers of clusters $k$. The insets show how $1-r$ depends on the number of clusters $k$ at different inverse temperatures indicated by the colors of the curves ($\beta=0.3,0.58,0.86,1.14,1.42,1.7$).
The numbers of samples $M$ are scaled with the numbers of clusters $M = k \times 5 \times 10^4$ to ensure a constant average number of states per cluster. The system size is $N=100$. {\bf a)} Reconstruction based on mean-field approximation. {\bf b)} Reconstruction based on the TAP approximation with gradient descent.
\label{fig_SK}}
\end{figure}

We test this approach using couplings drawn independently from a Gaussian distribution of zero mean and variance $1/N$ (the Sherrington--Kirkpatrick model~\cite{SK}). Figure~\ref{fig_SK}a shows the reconstruction at low temperatures improving with the number of clusters $k$ and configuration samples $M$. We note that at high temperatures, magnetisations are $0 \pm {\mathcal O}(N^{-1/2})$, so for small system sizes clustering erroneously identifies distinct clusters with small magnetisation. Thus at high temperatures, the low temperature resconstruction based on many clusters does not work as well as the standard approach based on a single cluster.

A further improvement is possible. For disordered systems, the self-consistent equation~\eqref{nMF_selfconsistent} is not exact. 
An additional term is required, the so-called Onsager reaction term describing the effect a spin has on itself via the response of its neighbouring spins. 
The Thouless--Anderson--Palmer (TAP) equation~\cite{TAP}
\begin{equation}
\label{TAP_selfconsistent}
m_i=\th ( \sum_{j \neq i} J_{ij} m_j - m_i \sum_{j \neq i} J_{ij}^2 (1-m_j^2)+ h_i) 
\end{equation}
turns out to be exact for models where all spins interact with each other. For each state $a$ we now obtain instead of~\eqref{MF_reconstruction_up}
\begin{eqnarray}
 \label{TAP_reconstruction} 
&&\sum_{j \neq i} J_{ij}  \bar{c}_{jk}^a=   - \delta_{ik} + \bar{c}_{ik}^a/(1-(\bar{m}_i^a)^2) + \nonumber \\
&&\ \  \bar{c}_{ik}^a \sum_{j \neq i} J_{ij}^2 (1-(\bar{m}_j^a)^2) - 2\bar{m}_i^a  \sum_{j \neq i} J_{ij}^2  \bar{m}_j^a \bar{c}_{jk}^a \ .
\end{eqnarray}
These equations are no longer linear in the couplings $J_{ij}$ and cannot be solved by the pseudoinverse. A simple gradient descent method, still allows to solve these equations 
in ${\mathcal O}(kN^3)$ steps per iteration. We define a quadratic cost function $S$ for the couplings ${\bf J}$ by squaring the difference between lhs and rhs of equation~\eqref{TAP_reconstruction}  and summing over all spin pairs $i,k$ and states $a$. 
Differences in the external fields $h_i^a=\arcth \bar{m}_i^a - \sum_{j \neq i} J_{ij} \bar{m}_j^a+ \bar{m}_i^a \sum_{j \neq i} J_{ij}^2 (1-(\bar{m}_j^a)^2)$ across thermodynamic states are penalized by an additional term 
$\sum_{i,a} \left( h_i^a - \langle h_i \rangle \right)^2$. The iterative prescription with rate $\eta$, $J_{ij} \leftarrow J_{ij} - \eta \partial S/\partial J_{ij}$, converges to a point near the solution of the TAP equation with small differences in the external fields across states (the deviations resulting from the finite number of samples and finite system size). 
Figure ~\ref{fig_SK}b shows how the reconstruction error asymptotically tends to zero with growing $k$ and $M$.

Mean-field theories exists beyond the Curie--Weiss, or the Sherrington--Kirkpatrick model discussed here~\cite{saad_opper_advancedMF}. We have shown that the use of mean-field methods to solve the inverse Ising problem at low temperatures hinges on our ability to reconstruct the thermodynamic states from the sampled data. With this proviso, the entire range of mean-field methods can be now be used, for instance for tree-like couplings~\cite{NguyenBerg} or couplings with local loops~\cite{kikutchi}.

We placed our focus on mean-field approaches, since they result in computationally efficient reconstructions independently of the underlying model (for instance 
a full connectivity matrix $J_{ij}$ versus a sparse matrix).  Reconstructions based on pseudo-likelihood~\cite{Ravikumar} can fail at low temperatures as well~\cite{bento2009}, although~\cite{pseudolikelihood} finds a good reconstruction for several models also at low temperatures, albeit at a large computational cost. 
The adaptive cluster expansion recently introduced by Cocco and Monasson~\cite{adaptive} is not affected by the transition to a low-temperature phase, but becomes computationally unwieldy for highly-connected models due to the large number of clusters to be considered.

{\bf Acknowledgements:} We thank Erik Aurell, Filippos Klironomos, and Nico Riedel for discussions. Funding by the DFG under SFB 680 is acknowledged.



\end{document}